# Growth of ultra thin vanadium dioxide thin films using magnetron sputtering

Fangfang Song[1] and B. E. White Jr.[1, a)]

*Department of Physics, Applied Physics, Astronomy, Binghamton University - State University of New York, Binghamton, NY, 13902*

In this work, the results of fabricating ultra thin $VO_2$ films on the technologically relevant amorphous $SiO_2$ surface using reactive DC magnetron sputtering are presented. Results indicate that a post deposition anneal in low partial pressures of oxygen is an effective way at stabilizing the $VO_2(M_1)$ phase on the $SiO_2$ surface. $VO_2$ films with a thickness of 42nm show a continuous microstructure, and undergo a resistivity change of more than a factor of 200 as the temperature of the film increases above $72^oC$. The film shows hysteresis in the metal-insulator transition temperature upon heating and cooling with a width of approximately $8^o$C. The resistivity of the low temperature semiconducting phase is found to be thermally activated with an activation energy $0.16\pm0.03$ $ev$. Stress measurements using X-ray diffraction indicate that the ultra thin $VO_2$ film has a large tensile stress of $2.0\pm0.2$ $GPa$. This value agrees well with the calculated thermal stress due to differential thermal expansion between the $VO_2$ thin film and silicon substrate. The stress leads to a shift of the metal-insulator transition temperature by approximately $4^oC$.

## I. INTRODUCTION

Vanadium dioxide is a well known strongly correlated material whose electronic properties have been actively investigated experimentally[1–6] and theoretically[7–12]. The interest in $VO_2$ results from the abrupt change in resistivity (over several orders of magnitude in bulk crystals) that occurs at a transition temperature near room temperature ($68^oC$). These features suggest numerous technological applications of $VO_2$, including novel memory devices[13,14], electronic switches[15,16], optical switches[17], and sensors[18].

However, the existence of a large number of distinct stable vanadium oxide phases offers a particular challenge to the growth of pure phase $VO_2$ thin films[19,20]. Extensive effort has been undertaken for the high quality thin film processing of this material. $VO_2$ thin films have been successfully prepared by various techniques, including sputtering[21–23], chemical vapor deposition[24–26], pulsed laser deposition[27–29] and sol-gel coating[30–32]. While the majority of the work in the literature has focused on growing thick $VO_2$ films (thickness > 100nm) under epitaxial growth conditions, relatively little has been reported on growing ultra thin vanadium oxide on an amorphous substrate. In addition, the literature indicates that growing high quality crystalline $VO_2$ films with thicknesses below 100nm is challenging. X. Wei argued that the crystallization of $VO_2$ deteriorates with decreasing film thickness and $VO_2$ films thinner than 100nm showed no obvious polycrystalline structure[33]. D. Brassard et al. also report the discontinuous nature of $VO_2$ thin films when the film thickness is below 50nm[34]. And Y. Park reported that a change of film thickness leads to a change of the crystalline phase and orientation of the vanadium oxide films[35].

Because the switching speed of many of the proposed $VO_2$ devices is proportional to the volume of $VO_2$ present, the ability to grow ultra thin films of the material while maintaining the metal-insulator transition is important. Incorporation of these new devices into exist-

a)ELectronic mail:bwhite@binghamton.edu

ing silicon based technology is made easier if these films can be produced on amorphous $SiO_2$ (a-$SiO_2$). In this paper, the investigation of the preparation of ultra thin $VO_2$ thin films on the technologically relevant $SiO_2$ surface by DC magnetron sputtering is presented. The results show that under proper deposition and post-annealing conditions, a factor of 200 change in resistivity can be achieved in films as thin as 42nm when deposited on an $SiO_2$ surface.

## II. EXPERIMENTAL

The vanadium oxide thin films were deposited using a sputtering system made by AJA International. Thin films were sputtered from a 99.9% pure vanadium metal target on a silicon substrate on which a 100Å thermal $SiO_2$ had been grown. The deposition chamber was evacuated to a pressure of $1.33 \times 10^{-4}$ Pa prior to heating the substrate. The deposition temperature was varied between $450°C$ to $550°C$ and the substrate was maintained at the desired temperature for 20 minutes for temperature stabilization. Pre-sputtering was performed for 10 minutes to clean the target. Reactively sputtered thin films were then deposited for a fixed deposition time of 10 minutes. During deposition, the pressure of the chamber was kept at 0.60 Pa. The Ar flow rate was set to be 20 $sccm$, and the $O_2$ flow rate was varied between 1.9 $sccm$ and 3 $sccm$ to vary the oxygen partial pressure. The DC power was maintained at 200 $W$, corresponding to a power density of 10.2 $W/cm^2$ across the 5 $cm$ diameter target. After deposition, the samples were annealed in an oxygen atmosphere. The oxygen pressure was varied from 0 Pa to $4.80 \times 10^{-2}$ Pa at an annealing temperature of $550°C$.

The crystallographic phase of the thin films was determined by Glancing Incident X-ray Diffraction(GIXRD). During GIXRD, the angle of incidence was fixed at $0.6°$, and the measurement was taken using a $2\theta$ scan geometry. The thickness of the sample was determined using X-ray Refelectivity. X-ray diffraction was also used to to measure the residual stress in the ultra thin $VO_2$ films. All diffraction scans were carried out using a PANalytical X'pert Pro Diffractometer with Cu $K_\alpha$ radiation. The microstructure of the thin films was investigated using a Zeiss Supra 55VP Field Emission Scanning Electron Microscope and a Veeco Multi-Techniques System 3100 Atomic Force Microscope with Nanoscope Voltage Controller. The electrical properties of the thin films were measured using a Keithley 4200 Semiconductor Characterization System.

## III. RESULTS AND DISCUSSION

### A. Effect of Deposition Conditions

#### 1. Effect of Oxygen Partial Pressure

The impact of oxygen partial pressure on the phase of the reactively sputtered thin films was determined by varying the percentage of $O_2$ in the sputtering gas from 9% to 11% while maintaining the total pressure in the sputtering chamber at 0.60 Pa. This small change in $O_2$ concentration produced profound changes in the phase of the thin films. Figure 1 shows the GIXRD results of $VO_x$ thin films deposited at slightly different oxygen partial pressure, while the substrate (thermally oxidized silicon) and deposition temperature ($550°C$) were held constant. When the percentage of oxygen in the sputtering ambient was 9%, two vanadium oxide phases were present in the films, $V_3O_5$ and $V_2O_3$; as the relative concentration of $O_2$ increased to 10%, the film was composed of $V_7O_{13}$, $VO_2$ and $V_2O_5$; a further increase in the relative oxygen concentration to 11% generates additional phases, including $V_2O_5$, $V_3O_7$, $V_6O_{13}$, $V_2O_3$, and $VO_2$. It can be seen that high oxygen partial pressure does not necessarily lead to a more oxygen rich phase.

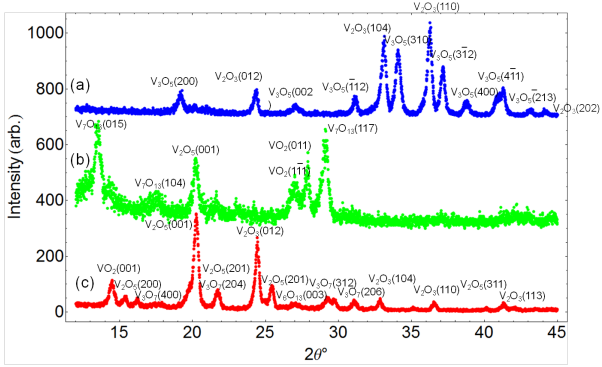

FIG. 1: GIXRD patterns of films prepared on same substrate (Si/SiO$_2$), same deposition temperature(550°$C$), but at different oxygen partial pressure,(a) 9% (b) 10% (c) 11%

### 2. Effect of Deposition Temperature

To investigate the impact of substrate temperature on the crystallographic morphology of the deposited films, the substrate temperature was varied from 450°C to 550°$C$ while maintaining the O$_2$ partial pressure at $6.00 \times 10^{-2}$ Pa. Figure 2 shows the GIXRD scans for the films prepared at different temperature. As shown in the figure, when the deposition temperature is 550°$C$, the film has two phases, V$_2$O$_3$ and V$_3$O$_5$; when the deposition temperature is 460°$C$, the film composition becomes V$_4$O$_9$,V$_3$O$_7$,V$_6$O$_{13}$ and V$_2$O$_5$; and when the deposition temperature is 450°$C$, the film has only one V$_2$O$_3$ phase. It can be seen that high substrate temperature does not necessarily produce an oxygen poor phase.

### 3. Effect of Substrate

To examine the impact of the substrate on the phases present in the thin films, two different substrates were explored, silicon and thermally oxidized silicon. Figure 3 shows the GIXRD pattern of samples prepared with the same deposition conditions but on different substrates (Si and oxidized silicon respectively). The deposition temperature was 450°$C$ and the oxygen partial pressure was $6.00 \times 10^{-2}$ Pa. It can be seen that the film is V$_2$O$_3$ on the oxidized silicon substrate, while VO$_2$(B) is formed on the Si substrate.

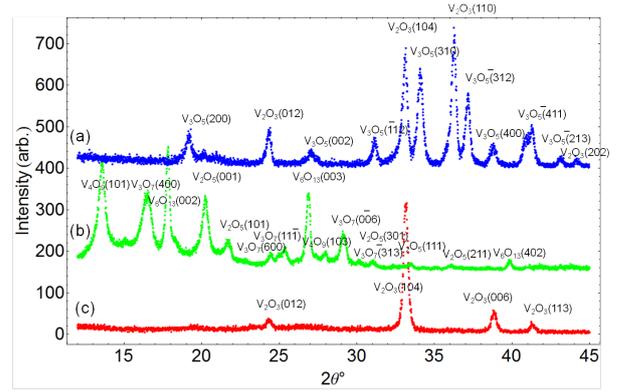

FIG. 2: GIXRD patterns of films prepared at same oxygen partial pressure(10%), on same substrate(Si/SiO$_2$), but at different temperature,(a) 550°$C$ (b)460°$C$ (c) 450°$C$

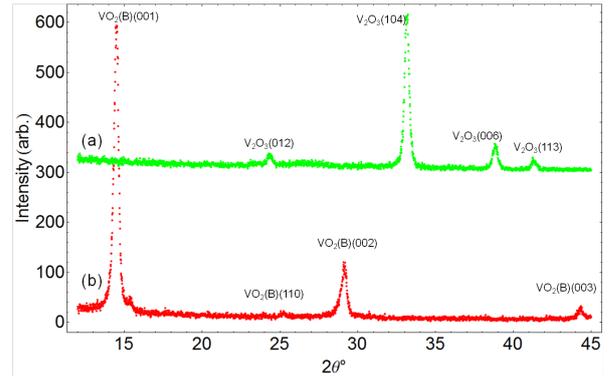

FIG. 3: GIXRD patterns of films prepared at same oxygen partial pressure($6.00 \times 10^{-2}$ Pa), same temperature(450°C), but different substrate, (a) Si/SiO$_2$ (b) Si

## B. Effect of Post Annealing Conditions

As shown in Figure 2, DC reactive sputtering under the explored conditions did not produce a pure $VO_2$ film on an amorphous $SiO_2$ surface. However, thin films of pure $V_2O_3$ were obtained on a-$SiO_2$ when the deposition temperature was $450°C$ with an oxygen partial pressure of $6.00 \times 10^{-2}$ Pa. Thus efforts were taken to convert the $V_2O_3$ phase into $VO_2$ by post-deposition annealing the sample in an oxygen atmosphere. The oxygen pressure was varied from 0 Pa to $4.80 \times 10^{-2}$ Pa at an annealing temperature of $550°C$

Figure 4 shows the GIXRD data for the vanadium oxide films after being annealed at $550°C$ for 50 minutes under different oxygen pressures. After the film is annealed at $2.00 \times 10^{-3}$ Pa of oxygen, all of the $V_2O_3$ peaks except for that of the (104) planes are observed to disappear. New peaks which are ascribed to the $V_3O_5$ phase appear. The peak at 30.20 ° corresponds to the $V_4O_7$ (122) plane. When the oxygen pressure becomes $6.70 \times 10^{-3}$ Pa, no $V_2O_3$ peaks are observed. Only the $V_4O_7$ (122) peak and peaks ascribed to $V_3O_5$ are present. When the oxygen pressure increases to $1.33 \times 10^{-2}$ Pa, peaks ascribed to the $VO_2(M1)$ phase appear. As the oxygen pressure is increased to $4.80 \times 10^{-2}$ Pa, all the peaks are ascribed to $VO_2$ indicating that a single phase thin film of $VO_2$ has been obtained.

Below, Figure 5 shows the GIXRD pattern of $VO_2$ films after being annealed at $550°C$ at an with oxygen partial pressure of $4.80 \times 10^{-2}$ Pa as a function of annealing time. After being annealed for 20 minutes, $V_2O_3$ peaks in the as deposited film are replaced by new peaks corresponding to the $VO_2(M_1)$ phase. As the annealing time increases, no new phases appear, and the intensity of peaks associated with the $VO_2(M_1)$ phase increases accompanied by decreases in the full width at half maximum (FWHM) of the peaks. This indicates grain growth

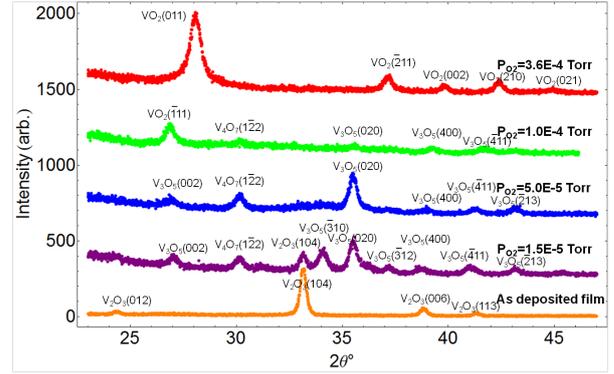

FIG. 4: GIXRD pattern of films after being annealed at different oxygen pressure

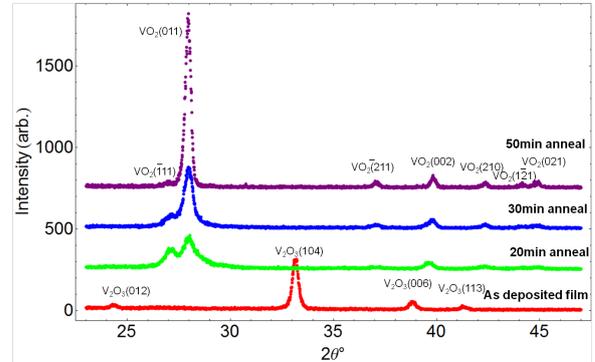

FIG. 5: GIXRD pattern of films after being annealed for different time

and an improvement of film crystallinity.

X-ray reflectivity was performed on the film after being annealed at $4.80 \times 10^{-2}$ Pa $O_2$ atmosphere for 50 minutes to determine the film thickness. As shown in Figure 6, the green line is the experimental data and the red line is the fitted data. The fitting parameters are summarized in table I. Based on these data, the thickness is determined to be 42nm.

### 1. Microstructure of The Thin Films

Figure 7 below shows SEM images of the film after different annealing times. In the as deposited film, the grains are uniform and densely packed with an average grain size of approximately 26.5nm . As the annealing time increases, grain growth is observed. Table II summarizes the average grain size as a function of annealing time.

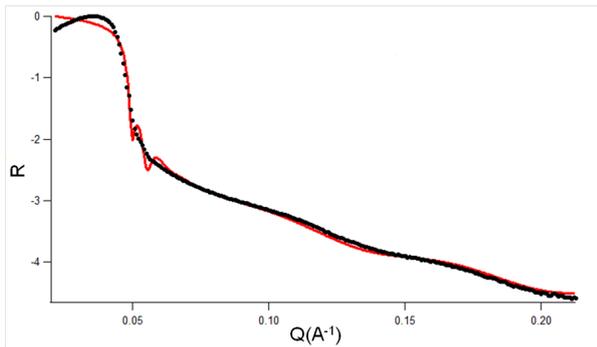

FIG. 6: X-ray reflectivity of the VO$_2$ film after being annealed at $4.80 \times 10^{-2}$ Pa atmosphere for 50 minutes

TABLE I: Refelectivity fitting parameter for VO$_2$ film

| Layer | Material | Thickness(Å) | Roughness(Å) | Re[SLD] (Å$^{-2}$) |
|---|---|---|---|---|
| Layer 1 | VO$_2$ | 422 | 55 | 4.5×10$^{-5}$ |
| Substrate | SiO$_2$ | - | 0.2 | 2.6×10$^{-5}$ |

TABLE II: Average grain size of vanadium oxide film after being annealed for different amount time

| Annealling time(min) | As deposited | 30 | 50 |
|---|---|---|---|
| Average Grain Size(nm) | 26.5 | 65.8 | 85.5 |

Figure 8 shows the AFM images of vanadium oxide films after being annealed for different amounts of time. And table III summarizes the Root Mean Square(RMS) roughness of these films. It can be seen that, going from as-deposited to being annealed for 30 minutes, the roughness of the film increases from 18.7Å to 54.4Å. After being annealed for 50 minutes, the roughness decreases to 43.5Å.

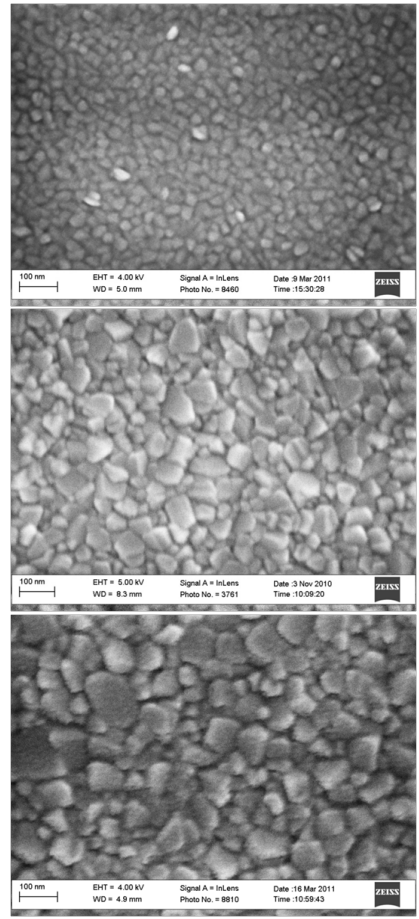

FIG. 7: SEM images of film after being annealed for different amount time (a) as deposited (b) 30 minutes (c) 50 minutes

TABLE III: RMS Roughness of vanadium oxide film after being annealed for different amount time

| Annealling time(min) | As deposited | 30 | 50 |
|---|---|---|---|
| RMS Roughness(Å) | 18.7 | 54.4 | 43.5 |

### 2. Temperature Dependence of Resistivity of The Thin Film

The nature of the hysteresis in the resistivity transition temperature of the VO$_2$ thin films after being annealed in $4.80 \times 10^{-2}$ Pa O$_2$ atmosphere at $550°C$ for 50 min-utes was examined using a Keithley 4200 Semiconductor Characterization System with a four probe collinear configuration. Sample temperature was controlled using a JANIS Research ST-500 Cryogenic Probe Station. The sample temperature was increased from room temperature to $127^oC$ followed by subsequent cooling back to room temperature.

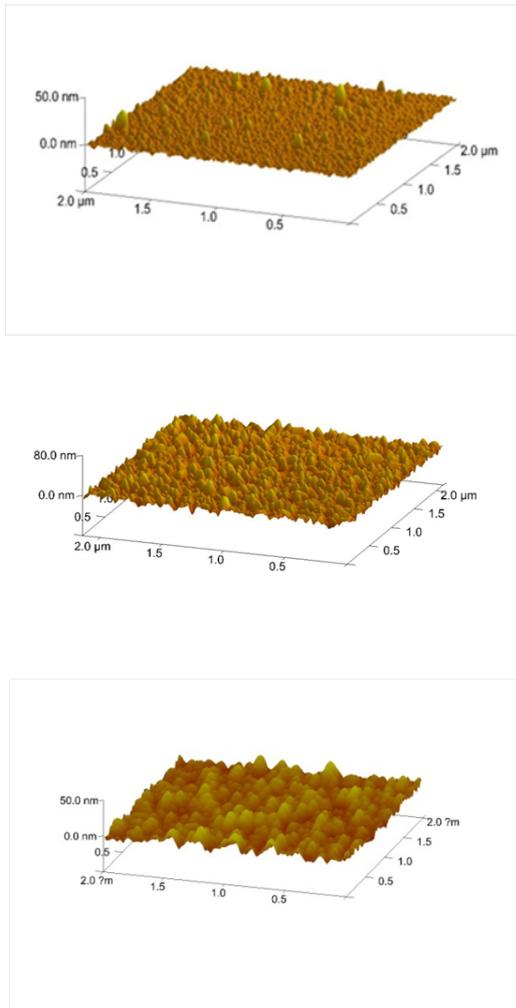

FIG. 8: AFM images of film after being annealed for different amount time (a) as deposited (b) 30 minutes (c) 50 minutes

The measured resistivity as a function of temperature is shown in figure 9. As can be seen in the figure, the transition temperature moves to higher temperatures when the transition temperature is approached from the high temperature side of the data. The width of the hysteresis is estimated to be approximately $8^\circ C$, which is similar to the hysteresis width in $VO_2$ films expitaxilly grown on sapphire approximately $6^\circ C$-$10^\circ C$.

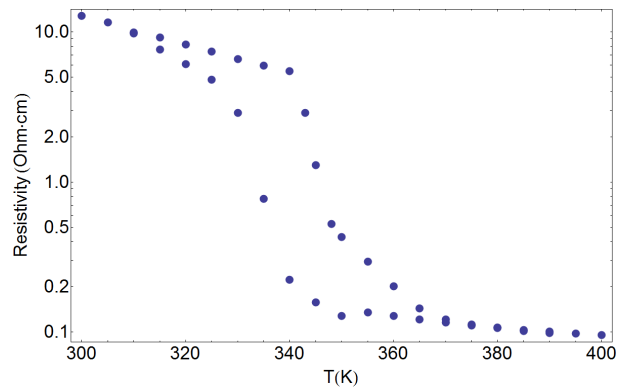

FIG. 9: Measured resistivity hysteresis curve of $VO_2(M_1)$ thin film

In order to precisely determine the transition temperature, we plot the $d(\log \rho)/dT$ vs. $T$ for the heating cycle and the temperature at the minimum point of the curve is chosen to be the transition temperature[36]. As shown in figure 10, the transition temperature is determined to be $72^oC$, which is higher than the reported value of $68^oC$ on a sapphire substrate.

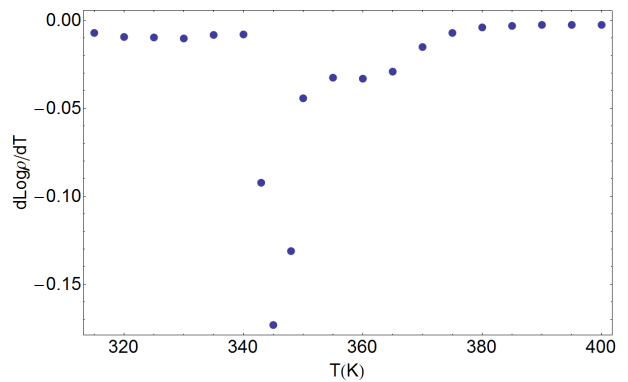

FIG. 10: $d(\log \rho)/dT$ vs. $T$ plot

As shown in figure 9, the resistivity of the low tem-perature monoclinic $VO_2$ phase on a a-$SiO_2$ substrate decreases as the temperature increases, which is a characteristic of a semiconductor. The temperature dependence of the resistivity has been observed to be thermally activated.

The activation energy for the conductivity is calculated using the Arrhenius equation:

$$\sigma = \sigma_a \exp(-E_a/kT) \quad (1)$$

Where $E_a$ is the activation energy, $\sigma$ is the conductivity of the film which is calculated using $\sigma = 1/\rho$, and $\sigma_a$ is the pre-exponent factor.

Figure 11 shows the plot of $\ln \sigma$ as a function of $1/T$ in the temperature range from $27^oC$ to $68^oC$. From the slope of the fitted line, the activation energy was calculated to be $0.16\pm0.03eV$, in agreement with the reported value of $0.188$ $ev$ for $VO_2$ films grown on a Si substrate[37] and $0.12$ $eV$ for $VO_2$ films grown on a sapphire substrate[38]. Although the temperature range is limited, Ref.[37] used a similar temperature range to calculate the activation energy.

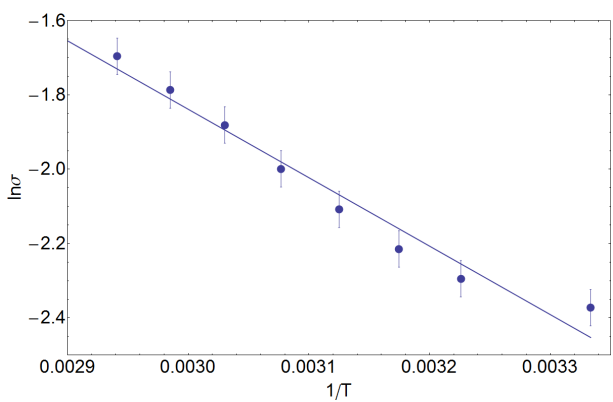

FIG. 11: $\ln \sigma$ vs. $1/T$ curve. The points are the measured data, and the solid line is the linear fitting line

## C. Measurement of Residual Stress in the Films

Stress is believed to play an important role in the temperature of the resistivity transition of $VO_2$[39–41]. Stress in the thin films of $VO_2$ was determined using the well-known $sin^2(\psi)$ method[42]. This method takes advantage of the fact that when a stress is applied to a polycrystalline material, the material will be deformed elastically, and the lattice plane spacing in the constituent grains will change from their stress free value to some new value corresponding to the applied stress. The change of plane spacing causes a shift in the angles at which the diffraction peaks are observed in XRD to new $2\theta$ positions. By measuring the $2\theta$ position at different sample tilt angles $\psi$, which is the angle between the diffraction plane normal and the thin film surface normal, the residual stress can be calculated as[42]:

$$\sigma_\phi = \frac{E}{(1+\nu)sin^2\psi}\left(\frac{d_i - d_n}{d_n}\right) \quad (2)$$

Where $d_n$ is the plane spacing at normal incidence under stress, $d_i$ is the plane spacing when the incident beam is inclined at angle $\psi$ to the surface normal, $E$ is the Young's modulus, and $\nu$ is the Possion ratio.

The slope of a linear plot between $\frac{d_i - d_n}{d_n}$ and $sin^2\psi$ equals $\frac{(1+\nu)\sigma_\phi}{E}$.

Although high $2\theta$ peaks are usually preferred for higher strain sensitivity, the higher $2\theta$ peaks are too weak and have shapes that are irregular, as shown in figure 5 for $VO_2$ films after being annealed at $4.80 \times 10^{-2}$ Pa $O_2$ for 50 minutes. Thus, the peak at $2\theta = 27.9^o$, which corresponds to diffraction from $VO_2$ (011) planes, was chosen for the stress measurement.

Figure 12 shows the XRD scan results with the incident beam inclined at different angles, $\psi$, to the surface normal. The grazing incident angle was set at $0.6^o$ and those measurements were taken in the $2\theta$ ranges from $26.5^o$ to $29.5^o$ using a $2\theta$ scan geometry. It can be seen that the 2θ peak shifts to a lower value as $\psi$ increases. Table IV below summarize the 2θ position and calculated lattice spacing at different $\psi$.

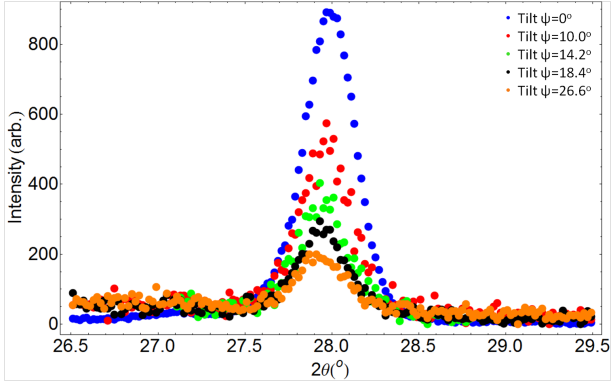

FIG. 12: GIXRD scan in range $26.5^o$-$29.5^o$ with different tilt angle $\psi$

TABLE IV: Summaried measured 2θ position and calculated lattice spacing at different tilt angle $\psi$

| $sin^2\psi$ | 0 | 0.03 | 0.06 | 0.1 | 0.2 |
|---|---|---|---|---|---|
| $\psi$ | 0 | 10.0 | 14.2 | 18.4 | 26.6 |
| 2θ | 27.99 | 27.97 | 27.95 | 27.94 | 27.89 |
| Lattice Spacing(d) | 3.1851 | 3.1871 | 3.1902 | 3.1902 | 3.1965 |

Figure 13 is the plot of $\frac{d_i-d_n}{d_n}$ as a function of $sin^2\psi$. The solid line is the best linear fit to the data. The slope of this line is found to be 0.018±0.002. Taking the VO$_2$ Young's modulus to be 140 $GPa$[43], and Possion ratios to be 0.3, the stress in the VO$_2$ thin film is calculated to be 2.0±0.2 $GPa$. The positive slope of the $\frac{d_i-d_n}{d_n}$ vs. $sin^2\psi$ plot indicates the stress is tensile.

As the thermal expansion coefficient of VO$_2$ is reported to be 2.1×10$^{-5}$/C[44], which is almost 10 times larger than that of Si, the film will contract more than the substrate when the film is cooled from 550°C to room temperature after annealing, producing a tensile stress in the film. Since the substrate is very thick compared to the film,

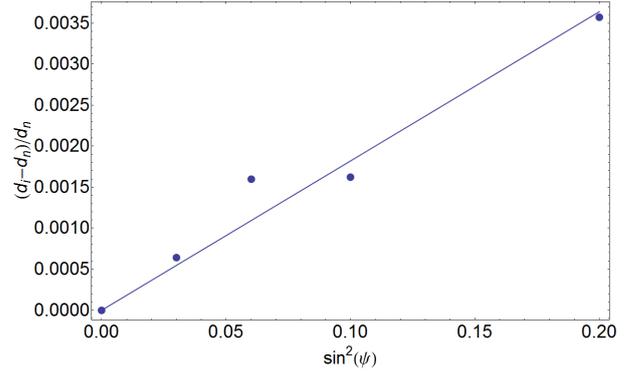

FIG. 13: Plot of $\frac{d_i-d_n}{d_n}$ as a function of $sin^2\psi$, the solid line is the best linear fitting line

the equation for an infinitely thick substrate[45] can be used to calculate the thermal stress in the film:

$$\sigma_T = \int_{T_0}^{T_f} \Delta\alpha \frac{E}{1-\nu}dT = \Delta\bar{\alpha}\frac{E}{1-\nu}\Delta T = \frac{E}{1-\nu}\Delta\bar{\alpha}\Delta T \quad (3)$$

If we assume the thermal expansion coefficient of the VO$_2$ film and substrate do not vary with temperature, the calculated stress due to differential thermal expansion is found to be:

$$\sigma_T = \frac{140 GPa}{1-0.3} \times (2.1 \times 10^{-5} - 2.7 \times 10^{-6}) \times (550-25)$$
$$= 1.92 GPa \quad (4)$$

This value agrees well the measured result, indicating that the residual stress in the VO$_2$ film mainly comes form thermal stress. In 2010, a stress-temperature phase diagram for the metal-insulator transition in VO$_2$ using a free-standing VO$_2$ nanobeam[39] was generated. This work indicates that the metal-inulator transition temperature should shift to higher values when the film is under tensile stress, as observed here.

## IV. CONCLUSION

Preparation of ultra thin $VO_2$ films on the technologically relevant amorphous $SiO_2$ surface has been studied. The effect of oxygen partial pressure, deposition temperature and substrate on the film composition and crystal structure have been discussed. Post annealing in at low pressures of $O_2$ iwas found to stabilize the VO2($M_1$) phase.

$VO_2$ films with a thickness of 42nm show a continuous microstructure and undergo a resistivity change of more than a factor of 200 as the temperature increases above the metal-insulator transition temperature. The transition temperature was determined to be $72^oC$. The film shows hysteresis in the transition temperature when heating and cooling with a width of approximately $8^oC$. The activation energy of the low temperature semiconducting phase of the $VO_2$ film is determined to be $0.16\pm0.03 ev$. X-ray diffraction was used to measure the residual stress in the $VO_2$ films. Results indicate that the ultra thin $VO_2$ film has a large tensile stress of $2.0\pm0.2$ $GPa$. This value agrees well with that calculated thermal stress assuming differential thermal expansion between the $VO_2$ film and substrate. The stress is expected to lead to a shift of the metal-insulator transition temperature, as observed in the temperature dependent resistivity measurement.

## V. ACKNOWLEDGEMENTS

We gratefully acknowledge the Center for Advanced Microelectronics Manufacturing at Binghamton University for support of this project. We are also indebted to Shijun Yu for assistance with the stress measurements in this work.